\documentclass[prl,twocolumn,showpacs,amsmath,amssymb,superscriptaddress]{revtex4-1}

\usepackage{setspace}

\usepackage{amsmath}
\usepackage{amsfonts}            
\usepackage{amssymb}

\usepackage{mathrsfs}

\usepackage{eufrak}
\usepackage{color}

\usepackage{graphicx}
\usepackage{epstopdf}

\usepackage{dcolumn}
\usepackage{bm}

\makeatletter
\def\@dotsep{4.5}
\makeatother

\begin{document}

\title{Non-adiabatic preparation of spin crystals with ultracold polar molecules}

\author{Mikhail Lemeshko}
\email{mikhail.lemeshko@gmail.com}
\affiliation{ITAMP, Harvard-Smithsonian Center for Astrophysics, 60 Garden Street, Cambridge, MA 02138, USA}%
\affiliation{Physics Department, Harvard University, 17 Oxford Street, Cambridge, MA 02138, USA} %

\author{Roman V. Krems}
\email{rkrems@chem.ubc.ca}
\affiliation{Department of Chemistry, University of British Columbia, Vancouver, BC V6T 1Z1, Canada }

\author{Hendrik Weimer}
\email{hweimer@cfa.harvard.edu}
\affiliation{ITAMP, Harvard-Smithsonian Center for Astrophysics, 60 Garden Street, Cambridge, MA 02138, USA}%
\affiliation{Physics Department, Harvard University, 17 Oxford Street, Cambridge, MA 02138, USA} %

\date{\today}

\begin{abstract}

We study the growth dynamics of ordered structures of strongly interacting polar molecules in optical lattices. Using dipole blockade of microwave excitations, we map the system onto an interacting spin-$1/2$ model possessing ground states with crystalline order, and describe a way to prepare these states by non-adiabatically driving the transitions between molecular rotational levels. The proposed technique bypasses the need to cross a phase transition and allows for the creation of ordered domains of considerably larger size compared to approaches relying on adiabatic preparation.  


\end{abstract}


\maketitle


Long-range dipolar interactions enable the creation of novel states of matter with ultracold quantum gases~\cite{LahayePfauRPP2009}. A prominent example is
the predicted formation of dipolar crystals with tunable interaction parameters
\cite{BuchlerPRL07,AstrakharchikPRL07,CapogrossoPRL10}.  However,
the preparation of such strongly interacting phases starting from a
weakly interacting quantum gas is known to be very challenging, as it
involves crossing a phase transition, where the energy gap vanishes in
the thermodynamic limit \cite{SachdevQPT}. In this Letter we show that for
ultracold polar molecules in an optical lattice, this problem can be
overcome by non-adiabatic driving of rotational transitions.

Our approach builds on recent experimental advances in the coherent
creation and control of ultracold polar alkali metal dimers, such as
KRb~\cite{NiScience08, *AikawaPRL10} and LiCs~\cite{DeiglmayrPRL08}.
The experiments have demonstrated that ultracold polar molecules 
can now be produced in the rovibrational ground
state,  transferred to any hyperfine
sublevel~\cite{OspekausPRL10}, and trapped in a periodic potential of an
optical lattice~\cite{MirandaYeJin11}. Ultracold molecules trapped on an optical lattice have previously been proposed as promising candidates for quantum computation \cite{DeMillePRL2002} and quantum
simulation of spin-lattice models~\cite{LahayePfauRPP2009,KreStwFrieColdMolecules,*BaranovPhysRep08,*TrefzgerJPB11,
GorshkovPRA11}.  Here, we employ dipole blockade of microwave excitations in the context of an effective spin-$1/2$ model with ultracold molecules on an optical lattice.  The ground state phase diagram of such model is dominated by a series of commensurate phases, in which one of the spin states exhibits
crystalline order. We show that these phases can be prepared 
by a short sequence of microwave pulses that nucleate the ordered domain, followed by continuous microwave driving that propagates the domain boundary. We
provide an effective model that describes the dynamics under
the continuous  driving and demonstrate that it leads
to an efficient growth of the ordered domains. Finally, we analyze the
imperfections and the required experimental parameters, and
demonstrate that structures consisting of $\gtrsim 1000$ spins can be grown.

\begin{figure}[b]
\includegraphics[width=\linewidth]{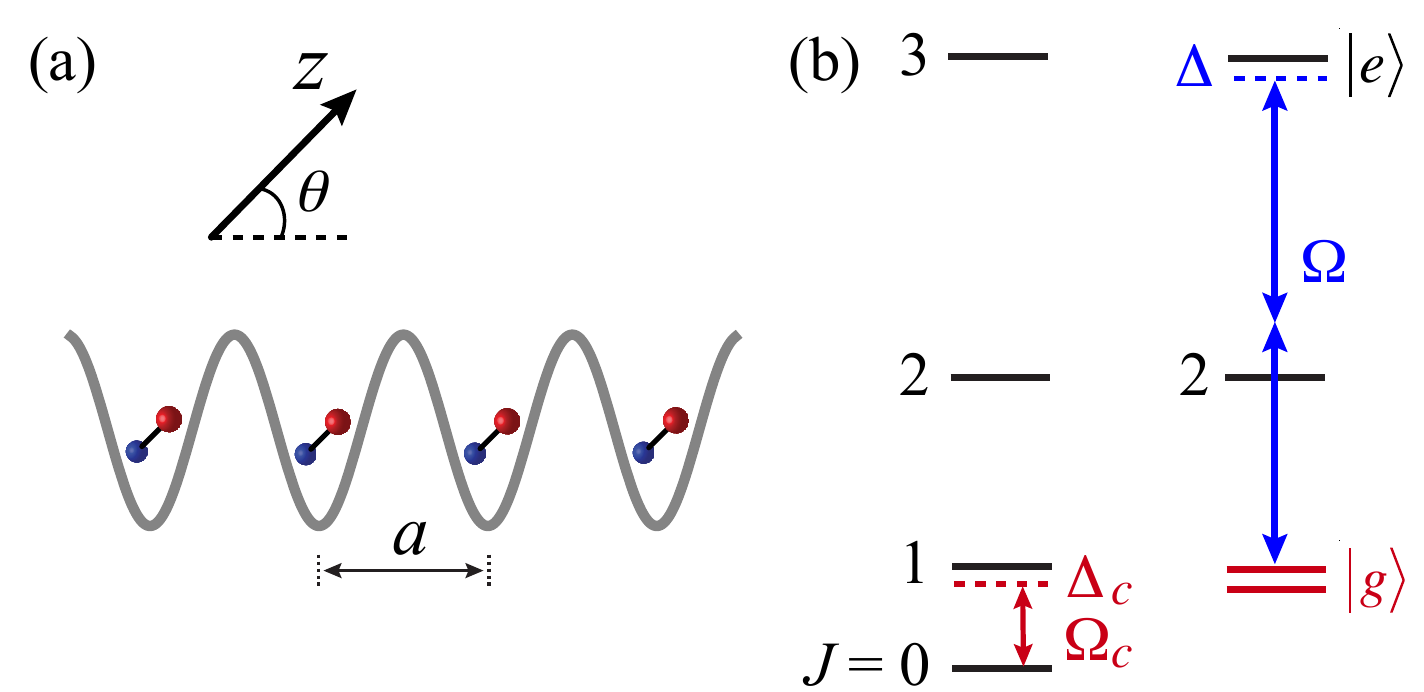}
\caption{\label{fig:levels} (a) 1D array of polar molecules on an optical lattice tilted by the angle
  $\theta$ with respect to the laboratory $z$-axis. (b) Lowest rotational states
  of a polar $^1\Sigma$ molecule. A strong microwave field
  $\Omega_{c}$ couples the states $\vert 0, 0\rangle$ and $\vert 1,
  0\rangle$, providing them with permanent dipole moments in the
  rotating frame; a weak field $\Omega \ll \Omega_{c}$ drives the
  two-photon transition between one of the dressed states $\vert g
  \rangle$ and the state $\vert e \rangle = \vert 3, 0 \rangle$, with the single-photon
  frequency far detuned from the $J=2$ state. All fields are linearly
  polarized along $z$.}
\end{figure}

To explore these ideas, we consider a one-dimensional  array of $^1\Sigma$ molecules with a dipole moment $d$ prepared in a Mott insulator state in an  optical lattice with the period $a$, tilted by the angle $\theta$ with respect to the laboratory $z$ axis, as shown in
Fig.~\ref{fig:levels}~(a). The lattice is filled with one molecule per
site and the trapping potential is strong enough to prevent tunneling between the sites and
molecular reactions~\cite{MirandaYeJin11}. The rotational states of a $^1\Sigma$
diatomic molecule, $\vert J, M \rangle$, are labeled by the angular
momentum $J$ and its projection on the laboratory $z$-axis, $M$. 
The energy of the rotational states is given by $E_\text{rot} = B J(J+1)$, where $B$ is the rotational constant.
We assume that the molecules are initially prepared in the rotational ground $J=0$ state.

The states $| J, M \rangle$ can be coupled by a microwave field, producing linear combinations of rotational states with different parity. In particular, the rotational states $\vert 0, 0\rangle$ and $\vert 1, 0\rangle$ can be coupled by near-resonant linearly polarized microwave field with Rabi frequency $\Omega_{c}$ and detuning $\Delta_{c}$ to produce 
two field-dressed states separated by $\sim \Omega_c$.
If the microwave field is applied adiabatically, the molecules must all populate the same field-dressed state. We choose this state, $\vert g \rangle = a \vert0,0\rangle
+ b \vert 1,0\rangle$, as the ground, `spin-down', state. 
In the rotating frame, the state $\vert g
\rangle$ has a permanent dipole moment $d_g = (\textcolor{black}{\sqrt{2}} a b/\sqrt{3}) d$,
leading to the dipole-dipole interaction between the molecules,
$V_\text{dd} (r) = d_g^2/r^3 (1-3 \cos^2 \theta)$, whose magnitude and
sign are tunable by changing $\Delta_c/\Omega_c$ and $\theta$
\footnote{In the laboratory frame the molecular dipole moments are
rapidly oscillating, which does not affect $V_\text{dd}$ assuming that
that the wavelength of the coupling field $\lambda_c \gg a$, which is
a good approximation for microwaves.}. We assume the coupling
field to satisfy the condition $V_\text{dd}(a) \ll \Omega_{c} \ll 2B$
that ensures that the dipole-dipole interaction does not mix $\vert g
\rangle$ with the neighboring field-dressed state nor with the dark $\vert
1, 1 \rangle$ state, which for typical experimentally realizible systems (molecules with dipole moment 1 -- 5 Debye and $a \sim 250$ -- $500$ nm) requires
$\Omega_{c} \sim 1-100$~MHz. As the effective spin-up state we choose
a level without a permanent dipole moment, $\vert e \rangle = \vert
3,0 \rangle$.  The two pseudo-spin states, $\vert g \rangle$ and $\vert
e \rangle$, are connected by a weak two-photon transition with
Rabi frequency $\Omega \ll \Omega_c$, with the single photon resonance
far detuned from the $\vert 2,0 \rangle$ state, which remains
unaffected by the microwave field.

Using the rotating wave approximation, we can write the Hamiltonian for an ensemble of molecules on an optical lattice as
\begin{multline}
\label{Hamil1}
	H= - \hbar \Delta  \sum_i \vert e_i \rangle \langle e_i \vert + \frac{\hbar \Omega}{2}  \sum_i ( \vert e_i \rangle  \langle g_i \vert + \vert g_i \rangle \langle e_i \vert) \\ +  V   \sum_{j<i} \frac{ \vert g_i, g_j \rangle \langle g_i, g_j \vert}{\vert i - j \vert^3},
\end{multline}
where $V = V_\text{dd}(a)$ and $\Delta = \omega - \omega_{ge}$ is the detuning of the two-photon field from the $ g  - e$ resonance. Eq.~(\ref{Hamil1}) can be expressed via spin-$1/2$ operators
$S_\alpha$ as
\begin{equation}
\label{Ising}
	H= h_z \sum_i S^{(i)}_z +  h_x  \sum_i S^{(i)}_x + V \sum_{j<i} \frac{S^{(i)}_z S^{(j)}_z}{\vert i - j \vert^3},
\end{equation}
where $h_x= \hbar \Omega$, $h_z = -(\hbar \Delta + \zeta(3) V)$, and
$\zeta(3) = \sum_{k=1}^\infty 1/k^3 \approx 1.202$ is Riemann's zeta
function \cite{RobicheauxPRA05}. The magnitude and sign of the
parameters $h_z/h_x$ and $V/h_x$ can be tuned by changing $\Delta$,
$\theta$, and $\Delta_c/\Omega_c$.  Note that the two-photon
transition driven by $\Omega$ is key to realizing the dipole
blockade~\cite{JakschPRL00, *LukinPRL01} with molecular rotational
levels: the pseudo-spin states $\vert g \rangle$ and $\vert e \rangle$
have a difference in angular momentum of $\Delta J \geq 2$, and
therefore are not mixed by the dipole-dipole interaction. This leads
to strong interactions between molecules in the $\vert g \rangle$
state, while eliminating the interactions between molecules in the $|g
\rangle$ and $|e\rangle$ states (the ``flip-flop'' terms).

The thermodynamic properties of the Hamiltonian (\ref{Ising}) have
been studied before in the context of Rydberg atoms
\cite{WeimerPRL08,WeimerPRL10,SelaPRB11}, and similar implementations
based on polar molecules have also been discussed previously
\cite{SchachenmayerNJP10}. The main advantage of using polar molecules
is that the dynamics occurs within the manifold of low-energy
rotational states, whose lifetimes ($>$ 1 s) are much longer than the lifetimes of Rydberg states ($\sim \mu$s). Therefore, 
polar molecules offer the possibility of creating ordered structures with much 
larger size.

\begin{figure}[t]\vspace{-0.4cm}
\includegraphics[width=\linewidth]{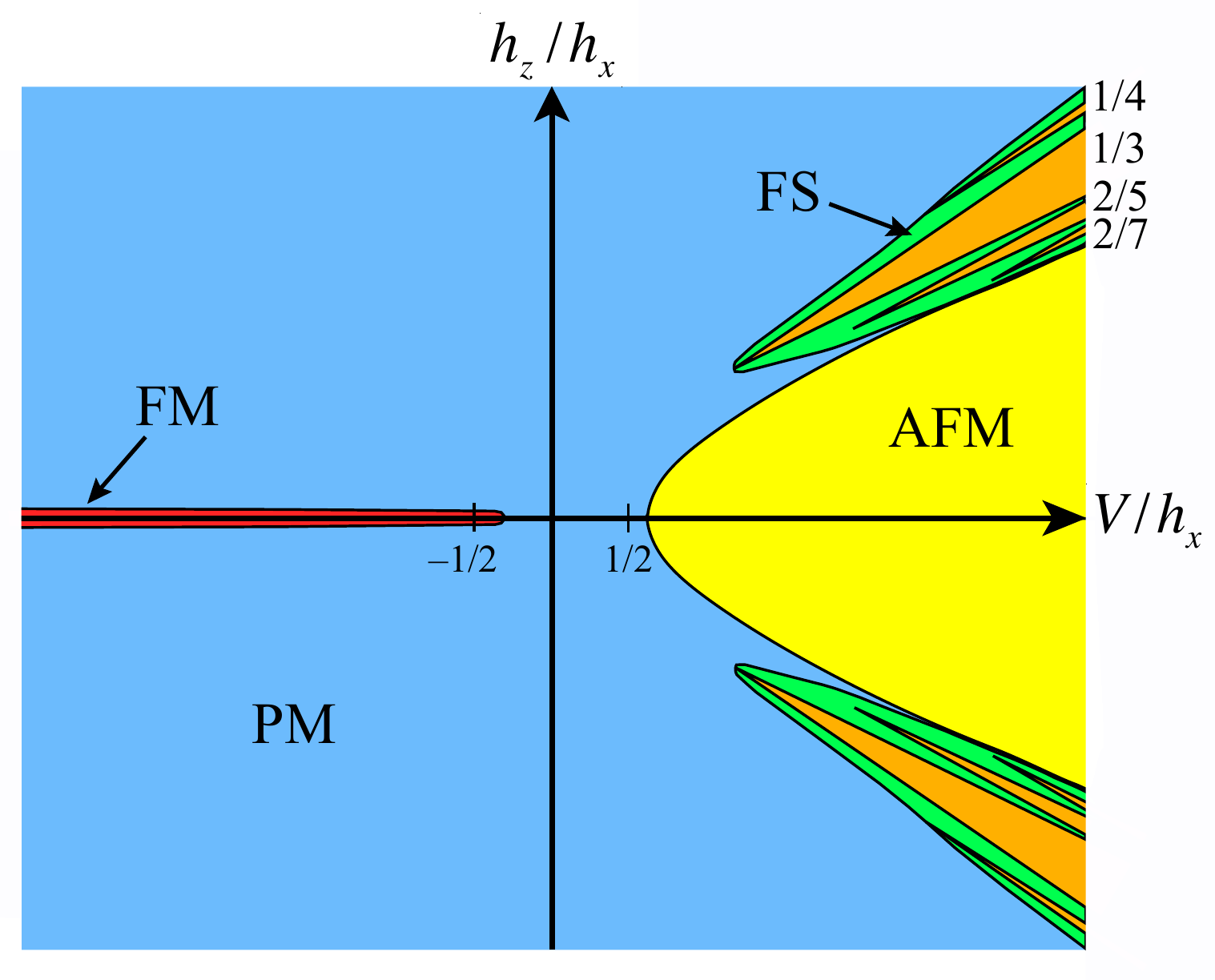}
\caption{\label{fig:phase_diagram} Schematics of the ground-state phase diagram corresponding to the Hamiltonian (\ref{Ising}). The commensurate crystal phases with indicated fillings (orange) are surrounded by the floating solid (FS) phase (green).}
\end{figure}

The main features of the phase diagram of the system described by
Eq.~(\ref{Ising}) are shown in Fig.~\ref{fig:phase_diagram} and can be
best understood in terms of the filling with which the minority spin
component occurs. When $V$ is positive and dominates over both $h_z$
and $h_x$, the system is in an antiferromagnetic (AFM) phase,
corresponding to half-filling, i.e. the up and down spins occur in an
alternating order. In the absence of a longitudinal field, $h_z =0$,
with $V$ being large and negative, all the spins are pointing in the
same direction, which corresponds to the ferromagnetic (FM)
ordering. There, any finite value of $h_z$ breaks the $\mathbb{Z}_2$
symmetry of the model leading to the breakdown of the FM phase
\cite{UzelacPRB80}. Note that the long-range interaction leads to the
opposite shifts of the transition points on the $h_z=0$ line. In the
FM phase, the long-range terms lead to renormalization of the nearest
neighbor coupling constant, while for the AFM interaction they result
in frustration accelerating the melting of the phase.  From both sides
of the AFM phase there is a complete devil's staircase of crystalline
configurations with different commensurate lattice
spacings~\cite{BakPRL82, *BurnellPRB09, SelaPRB11,
WeimerPRL10}. \textcolor{black}{In general, the commensurate phases melt
in two steps: first into the floating solid (FS) phase, and then into
the paramagnet (PM). The FS phase is characterized by a finite density
of Bose condensed dislocation defects of the commensurate crystal and
exhibits a gapless excitation spectrum \cite{WeimerPRL10}. At the tip
of the lobe at filling $1/3$ and $1/4$, however, the system belongs to
the Potts universality class, which results in a direct transition
from the commensurate phase to the PM \cite{SelaPRB11}.}

Adiabatic preparation of the ordered phases, as e.g.\ proposed in
Ref.~\cite{SchachenmayerNJP10}, is challenging due to the vanishing
energy gap between phases on the phase boundary. For a system of $N$
particles the gap closes as $1/N$ for the Ising transition between the
PM and the AFM phases \cite{ZurekPRL05} and for the transition between
the PM and the FS phases \cite{WeimerPRL12}, and as $1/N^2$ for the
transition between the FS phase and the commensurate crystals due to
its relation to free fermions \cite{WeimerPRL10,SachdevQPT}.  Within
the Landau-Zener approximation \textcolor{black}{(equivalent to the
Kibble-Zurek formalism for continuous phase transitions)} for the
adiabatic crossing of the phase boundary \cite{ZurekPRL05}, the
average size of the ordered domains is proportional to the square root
of the sweep duration. While, in some cases, this scaling can be
improved using non-linear sweeps \cite{QuanNJP10,WeimerPRL12}, it is
highly desirable to have a more robust protocol. Here, we present a
non-adiabatic method for creating crystalline order in the system,
with the growth of the domains scaling much better than within the
adiabatic approach.



\begin{figure}[t]
\includegraphics[width=\linewidth]{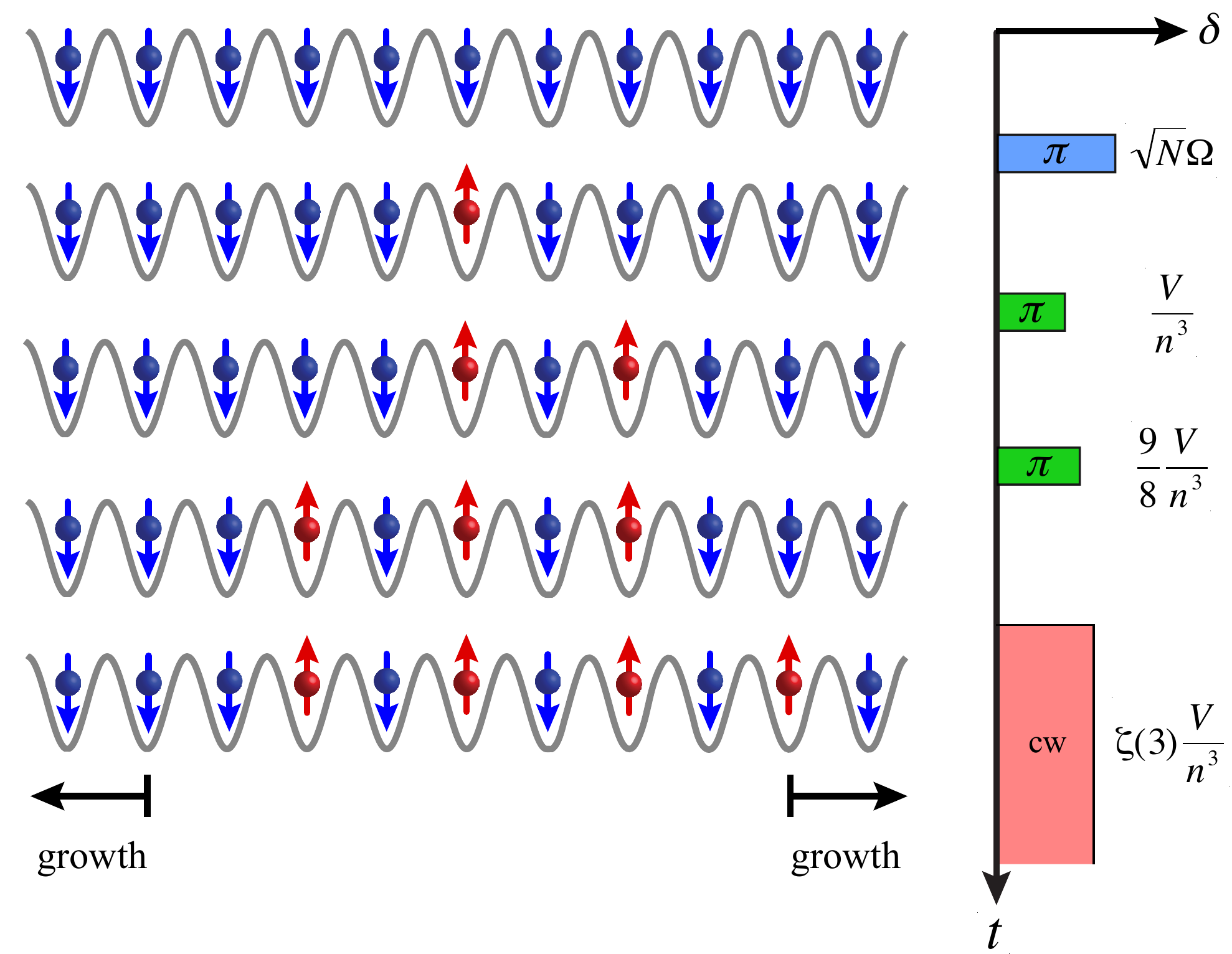}
\caption{\label{fig:pulses} Pulse sequence used to non-adiabatically
  prepare an ordered phase with a filling $1/n$: (i) a far detuned
  $\pi$ pulse (blue) nucleates the phase by flipping one spin per $N$
  lattice sites; (ii) a few resonant $\pi$-pulses (green) deterministically
  create a few lattice cells of the phase near the nucleation center;
  (iii) continuous driving (red) propagates the crystal boundary. The
  required values for the renormalized detunings of the pulses,
  $\delta=\Delta - E_1$, are shown. The lattice plots exemplify the
  $n=2$ (AFM) case.}
\end{figure}

In the following, we use the dipole blockade of microwave excitations to non-adiabatically construct strongly interacting ground state phases of the Ising model, eq.~(\ref{Ising}), as schematically shown in Fig.~\ref{fig:pulses}. We consider the case of $V>0$, corresponding to $\theta=\pi/2$, and exemplify the technique by caclulations for $^7$Li$^{133}$Cs molecules ($d=5.520$~Debye, $B=5.636$~GHz~\cite{AymarDulieuJCP05}) in an optical lattice with $a=266$~nm.

The procedure to grow large ordered domains is schematically illustrated in
Fig.~\ref{fig:pulses}. In the absence of the spin-flipping field $\Omega$, all the molecules are initialized in the ground (spin-down) state $| g \rangle$. 
The energy needed to resonantly flip the first spin in the lattice is
given by $E_1 = - 2V\zeta(3)$, and we use it to renormalize the
detuning as $\delta = \Delta - E_1$. In the first step, we nucleate the
dipolar crystal using a $\pi$-pulse far detuned from $E_1$, which
corresponds to $\delta/\Omega = \sqrt{N}$, where $N$ is the number of
sites of the ordered structure that one wishes to prepare. Then, each
of the up spins separated by $\sim N$ sites serves as a center around
which the ordered phase is to be grown.

Next, we make use of the dipole-dipole interactions between the
molecules to deterministically grow the commensurate dipolar crystal
with filling $1/n$ near the nucleation center. In order to resonantly
flip a spin located $n$ sites away from the nucleation center we use a
$\pi$ pulse with detuning $\delta = V/n^3$, while frequency flipping
the third spin in the chain is shifted by $\delta = V/n^3 + V/(2n)^3 =
9V/(8n^3)$. Such detuned pulses will only affect these two spins,
flipping any other spin in the system will be highly off-resonant.
The size of the ordered structure that one can deterministically
prepare with $\pi$ pulses, $\Delta x_0$, is limited by the Rabi
frequency $\Omega$, which must be smaller than the
change of the detuning $\delta$ from one pulse to another. Assuming
$\Omega \sim 2\pi \times \textcolor{black}{50}$~Hz and LiCs molecules, whose
interaction is given by $V/\hbar = 2\pi \times \textcolor{black}{40.7}$~KHz, one can
deterministically prepare the states with $\Delta x_0 = 8 a$ for the
$n=2$ (AFM) phase and $\Delta x_0 = 9a$ for the $n=3$ phase.

\begin{figure}[t]
\includegraphics[width=\linewidth]{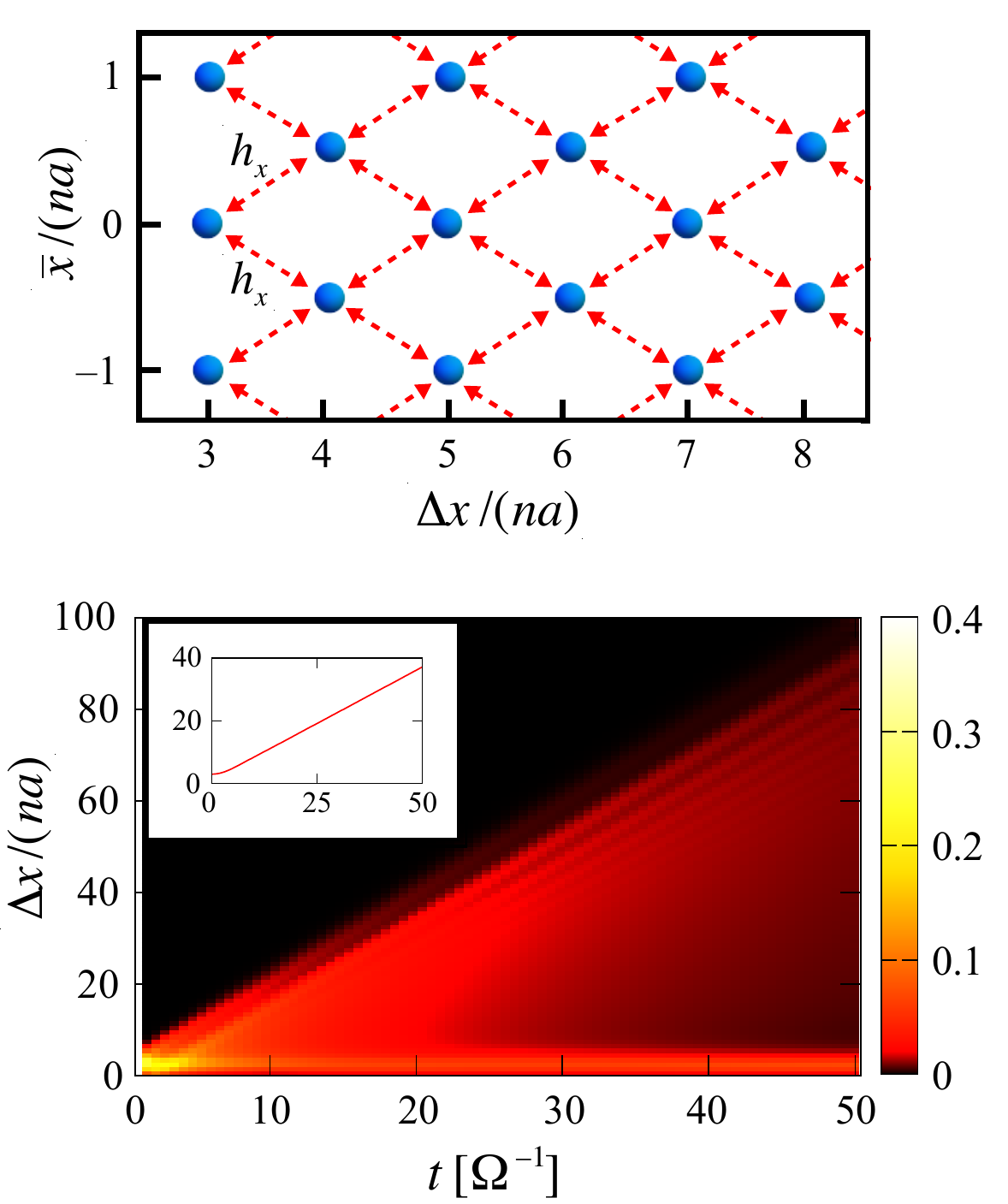}
\caption{\label{fig:walk} Top: the growth of the crystalline phase can
  be visualized as a particle hopping on a 2D lattice, with $\Delta x$
  the domain size, and $\bar{x}$ the domain center-of-mass position,
  as given by the effective Hamiltonian,
  eq.~(\ref{2DlatticeHamil}). The example is given for $\Delta x_0 =
  3na$. Bottom: the dynamics of the growth of the ordered phase during
  the continuous driving. Colors indicate the populations
  corresponding to different domain sizes, the inset
  \textcolor{black}{(same axis labels as in the main figure)} shows the
  root-mean-square of the size, which is linearly increasing with
  time.}
\end{figure}

After a few pulses the value of detuning required to flip the next
spin in the ordered chain approaches the value of $\delta = \zeta(3)
V/(na)^3$, and one can use continuous driving to grow the ordered
structure further. Then, the dynamics of the crystal growth is
equivalent to a single particle hopping on a semi-infinite two-dimensional
lattice $\{ i, j\}$ given by the domain size, $\Delta x_i \ge \Delta
x_0$, and the domain center of mass position, $\bar{x}_i$, as given by
the effective Hamiltonian,
\begin{multline}
\label{2DlatticeHamil}
	H_\text{eff} = \frac{\hbar \Omega}{2} \sum_{i,j}  \Bigl( \vert i,j \rangle \langle i+ 1, j\pm 1 \vert +  \vert i,j \rangle \langle i - 1, j\pm 1 \vert \Bigr) \\ + V\underset{k<i}{\sum_{i,j}} \frac{\vert i,j \rangle \langle i, j \vert}{(k n)^3},
\end{multline}
which is schematically shown in Fig.~\ref{fig:walk}
(a). \textcolor{black}{Numerically solving the time-dependent
Schr\"odinger equation by exact diagonalization, we study the dynamics
of the crystalline domain. As shown in Fig.~\ref{fig:walk}~(b), the
domain size increases linearly with time, illustrating that the growth process
is highly efficient.}

In order to prepare commensurate phases of higher order, such as $2/5$
or $2/7$, one can replace the continuous driving with a series of
pulses, $\{\pi_1, \pi_2, \pi_1, \pi_2 \dots \}$, with detunings
$\delta_1$ and $\delta_2$. In the case of the $2/5$ phase, whose unit
cell is given by $\uparrow \downarrow \downarrow \uparrow \downarrow$,
the required detunings are $\delta_1\approx 0.05~V$ and
$\delta_2\approx 0.14~V$. The frequency resolution needs to fulfill
the condition, $\Omega < \delta_2 - \delta_1$, which can be
easily satisfied for LiCs, requiring $\Omega \lesssim \textcolor{black}{4}$~KHz.

The leading source of errors in the process of growing the dipolar
crystal will be due to defects created by flipping a spin $n+1$ sites
(instead of $n$ sites) away from the domain boundary. During the
continuous driving, these states are detuned by $\Delta_{n+1} =
V(1+[\zeta(3)+\psi^{(2)}(1/n)/2]/n^3)$, where $\psi^{(n)}(z)$ is the
polygamma function \cite{AbramowitzStegun}. The number of spins that
can be flipped before creating a defect can be estimated by the
condition $N\Omega^2/\Delta_{n+1}^2 \approx 1$, which effectively
limits the Rabi frequency. Since the number of spins one can flip
within the lifetime $\tau$ of the system is given by
$N\approx\Omega\tau$, we can calculate the average size of the domains
to be $N=(\Delta_{n+1}\tau)^{2/3}$, \textcolor{black}{which scales more favorably with time than
the limit set by the Kibble-Zurek prediction.} Assuming $V/\hbar=2\pi\times \textcolor{black}{41}\,\mathrm{kHz}$
and the lifetime of the system of $\tau =
1\,\mathrm{s}$, we find that it is possible to prepare
domains of a $n=2$ crystal consisting of up to $nN = \textcolor{black}{1700}$ spins,
while for the $n=3$ crystal one can reach the size of $nN = \textcolor{black}{1000}$
spins. The value of $N$ determines the correlation length of the
defects, i.e.\ the effective temperature of the system, $T_\text{eff}
\sim \Delta_{n+1}/N $. Here, we find $T_\text{eff} =\textcolor{black}{2}00$~pK for LiCs
molecules and $N=1000$.

We note that in principle the performance of this procedure can be
improved even further, by using a composite pulse sequence that
dynamically decouples the processes leading to the creation of defects
\cite{DuerNMR}. Likewise, employing optical superlattice techniques
\cite{CheinetPRL08} could also be used to improve the addressing of
individual transitions.

Finally, we would like to emphasize that the dipolar crystals prepared after switching off the microwave driving correspond to
the ground state of the Hamiltonian (\ref{Ising}) in the limit of $h_x =
0$, and it is possible to explore the region of finite $h_x$ by
adiabatically turning the microwave driving field on again. Thus, it becomes possible to explore static and dynamic properties of
the commensurate phases. In particular, the dynamics of fractionalized
low-energy excitations can be studied, which are eventually
responsible for the transition to the floating solid phase
\cite{WeimerPRL10}.





\textcolor{black}{In summary, we have proposed a novel method to create
strongly interacting many-body states in the dipole blockade regime of
microwave transitions between rotational states of polar
molecules. Our approach relies on the non-adiabatic driving of the
microwave transitions and allows for the possibility to construct
larger domains of dipolar crystals than in an adiabatic scenario. The
scheme presented here is general and can be implemented with any two-state system possessing long-range Ising interactions, such as different
spin states of $^2\Sigma$ or $^2\Pi$ molecules, trapped ions
\cite{PorrasPRL04}, Rydberg atoms, or Nitrogen-Vacancy centers in
diamond \cite{WeimerPRL12}. Furthermore, our proposed method is not
limited to dipolar crystals and could be applied to other systems
where the initial state  can be efficiently coupled to the  many-body state of interest by
non-adiabatic driving.}


We thank Susanne Yelin, Samuel Meek, Nicolas Vanhaecke, and Charles Mathy for
insightful discussions. This work was supported by the National
Science Foundation through a grant for the Institute for Theoretical
Atomic, Molecular and Optical Physics at Harvard University and
Smithsonian Astrophysical Observatory and a fellowship within the
Postdoc Program of the German Academic Exchange Service (DAAD).




\bibliography{LemKremsWeim_Many-Body}
\end{document}